\documentclass[10pt, twocolumn]{IEEEtran}
\usepackage{amsthm}
\usepackage{amsmath}
\usepackage{amsfonts}
\usepackage{graphicx}
\usepackage{latexsym}
\usepackage{amssymb}
\usepackage{stmaryrd}
\usepackage{comment}
\usepackage{amscd}
\usepackage[small]{caption}
\usepackage[dvipsnames]{xcolor}
\usepackage{algorithm}
\usepackage{setspace}
\usepackage[noend]{algpseudocode}
\algrenewcommand\alglinenumber[1]{\scriptsize #1:}
\algrenewcommand\algorithmicindent{1em}%
\allowdisplaybreaks
\usepackage{graphicx}
\usepackage{subfigure}
\usepackage{wrapfig}
\usepackage{float}
\usepackage{bbm}
\usepackage{amsmath,bm}
\usepackage{mathrsfs}
\usepackage{tikz}
\tikzset{
    cross/.pic = {
    \draw[rotate = 45] (-#1,0) -- (#1,0);
    \draw[rotate = 45] (0,-#1) -- (0, #1);
    }
}
\usetikzlibrary{patterns}
\usetikzlibrary{decorations.pathreplacing,calligraphy}
\usepackage{graphicx} % Allows including images
\usepackage{booktabs} % Allows the use of \toprule, \midrule and \bottomrule in tables

\allowdisplaybreaks
\usepackage{mathtools}
\newcommand{\bea}{\begin{eqnarray}}
\newcommand{\eea}{\end{eqnarray}}
\newcommand{\bean}{\begin{eqnarray*}}
\newcommand{\eean}{\end{eqnarray*}}

\newcommand{\floor}[1]{\left\lfloor #1 \right\rfloor}

\newcommand{\sbinom}[2]{\left( \begin{array}{c} #1 \\ #2 \end{array} \right) }

\newcommand{\AoI}{\mathsf{AoI}}
\newcommand{\GE}{\mathsf{GE}}
\newcommand{\PEC}{\mathsf{PEC}}

\newcommand{\MDS}{\mathsf{MDS}}
\newcommand{\BEP}{\mathsf{BEP}}
\newcommand{\pers}{\cP}

\newcommand{\cA}{{\cal A}}
\newcommand{\cB}{{\cal B}}
\newcommand{\cC}{{\cal C}}

\newcommand{\cN}{{\cal N}}
\newcommand{\cO}{{\cal O}}
\newcommand{\cP}{{\cal P}}

\newcommand{\sG}{\script{G}}

\newcommand{\sP}{\script{P}}

\DeclareMathOperator*{\argmin}{arg\,min}

\DeclareMathAlphabet{\mathbfsl}{OT1}{cmr}{bx}{it}
\newcommand{\uuu}{\kern-1pt\mathbfsl{u}\kern-0.5pt}
\newcommand{\vvv}{\kern-1pt\mathbfsl{v}\kern-0.5pt}

\newcommand{\myboxplus}{\kern1pt\mbox{\small$\boxplus$}}

\makeatletter \DeclareRobustCommand{\sbinom}{\genfrac[]\z@{}}
\makeatother
\newcommand{\G}[2]{\sbinom{{#1}\kern-1pt}{{#2}\kern-1pt}}
\newcommand{\Gq}[2]{\sbinom{{#1}\kern-0.25pt}{{#2}\kern-0.25pt}}

\newcommand{\Ps}{\smash{{\sP\kern-2.0pt}_q\kern-0.5pt(n)}}
\newcommand{\sPs}{\smash{{\sP\kern-1.5pt}_q(n)}}
\newcommand{\Ptwo}{\smash{{\sP\kern-2.0pt}_2\kern-0.5pt(n)}}
\newcommand{\Ptwom}{\smash{{\sP\kern-2.0pt}_2\kern-0.5pt(m)}}
\newcommand{\Ptwonm}{\smash{{\sP\kern-2.0pt}_2\kern-0.5pt(n+m)}}
\newcommand{\Ptwoa}{\smash{{\sP\kern-2.0pt}_2\kern-0.5pt(1)}}
\newcommand{\Ptwob}{\smash{{\sP\kern-2.0pt}_2\kern-0.5pt(2)}}
\newcommand{\Ptwoc}{\smash{{\sP\kern-2.0pt}_2\kern-0.5pt(3)}}
\newcommand{\Ptwod}{\smash{{\sP\kern-2.0pt}_2\kern-0.5pt(4)}}
\newcommand{\Ptwoe}{\smash{{\sP\kern-2.0pt}_2\kern-0.5pt(5)}}
\newcommand{\Ptwof}{\smash{{\sP\kern-2.0pt}_2\kern-0.5pt(6)}}
\newcommand{\Ptwokm}{\smash{{\sP\kern-2.0pt}_2\kern-0.5pt(2k-1)}}
\newcommand{\Pone}{\smash{{\sP\kern-2.5pt}_2\kern-0.5pt(n{-}1)}}

\newcommand{\Gr}{\smash{{\sG\kern-1.5pt}_q\kern-0.5pt(n,k)}}
\newcommand{\Gi}{\smash{{\sG\kern-1.5pt}_q\kern-0.5pt(n,i)}}
\newcommand{\Gj}{\smash{{\sG\kern-1.5pt}_q\kern-0.5pt(n,j)}}
\newcommand{\Grmk}{\smash{{\sG\kern-1.5pt}_q\kern-0.5pt(n,n-k)}}
\newcommand{\Grdk}{\smash{{\sG\kern-1.5pt}_q\kern-0.5pt(2k,k)}}
\newcommand{\Grekappa}{\smash{{\sG\kern-1.5pt}_q\kern-0.5pt(n,e+1-\kappa)}}
\newcommand{\Grtwoekappa}{\smash{{\sG\kern-1.5pt}_q\kern-0.5pt(n,2e+1-\kappa)}}
\newcommand{\Gremkappa}{\smash{{\sG\kern-1.5pt}_q\kern-0.5pt(n,e-\kappa)}}
\newcommand{\Gn}{\smash{{\sG\kern-1.5pt}_2\kern-0.5pt(n,n{-}1)}}
\newcommand{\Gnq}{\smash{{\sG\kern-1.5pt}_q\kern-0.5pt(n,n{-}1)}}
\newcommand{\Gone}{\smash{{\sG\kern-1.5pt}_2\kern-0.5pt(n,1)}}
\newcommand{\Gqone}{\smash{{\sG\kern-1.5pt}_q\kern-0.5pt(n,1)}}
\newcommand{\GTwo}{\smash{{\sG\kern-1.5pt}_2\kern-0.5pt(n,k)}}
\newcommand{\GTwonk}[2]{{\smash{{\sG\kern-1.5pt}_2\kern-0.5pt({#1},{#2})}}}
\newcommand{\Gnk}{\smash{{\sG\kern-1.5pt}_2\kern-0.5pt(n,n{-}k)}}
\newcommand{\Greone}{\smash{{\sG\kern-1.5pt}_q\kern-0.5pt(n,e{+}1)}}
\newcommand{\Gretwo}{\smash{{\sG\kern-1.5pt}_q\kern-0.5pt(n,e{+}2)}}

\newcommand{\be}[1]{\begin{equation}\label{#1}}
\newcommand{\ee}{\end{equation}}

\newcommand\smallO{
  \mathchoice
    {{\scriptstyle\mathcal{O}}}% \displaystyle
    {{\scriptstyle\mathcal{O}}}% \textstyle
    {{\scriptscriptstyle\mathcal{O}}}% \scriptstyle
    {\scalebox{.7}{$\scriptscriptstyle\mathcal{O}$}}%\scriptscriptstyle
  }

\newcommand{\Cref}[1]{Co\-rol\-la\-ry\,\ref{#1}}

\newtheorem{theorem}{Theorem}
\newtheorem{lemma}{Lemma}

\newtheorem{corollary}{Corollary}

\newcommand{\B}{\boldsymbol}

\newcommand\blfootnote[1]{%
  \begingroup
  \renewcommand\thefootnote{}\footnote{#1}%
  \addtocounter{footnote}{-1}%
  \endgroup
}

\begin{document}

%  \author{%
%   \IEEEauthorblockN{\textbf{Shubhransh~Singhvi}\IEEEauthorrefmark{1}, 
%                      \textbf{Omer~Sabary}\IEEEauthorrefmark{2}, 
%                      \textbf{Daniella~Bar-Lev}\IEEEauthorrefmark{3}
%                      and \textbf{Eitan~Yaakobi}\IEEEauthorrefmark{3}}
%   \IEEEauthorblockA{\IEEEauthorrefmark{1}%
%                       Signal Processing  \&  Communications Research  Center, International Institute  of  Information Technology, Hyderabad, India}
%   \IEEEauthorblockA{\IEEEauthorrefmark{2}%
%                      Department of Electrical and Computer Engineering, %\\
%                      University of California San Diego, La Jolla, CA 92093, USA     }
%                         \IEEEauthorblockA{\IEEEauthorrefmark{3}%
%                      Department of Computer Science, %\\
%                      Technion---Israel Institute of Technology, 
%                      Haifa 3200003, Israel \vspace{-3.5ex}}
%  }

\title{Coding Gain for Age of Information in a Multi-source System with Erasure Channel}
\author{Shubhransh Singhvi  and Praful D. Mankar}
\date{\today}
\maketitle
\thispagestyle{empty}	
\pagestyle{empty}
%%%%%%%%
\begin{abstract}
In our work, we study the age of information ($\AoI$) in a multi-source system where $K$ sources  transmit updates of their time-varying processes via a common-aggregator node to a destination node through a channel with packet delivery errors. 
We analyze $\AoI$ for an $(\alpha, \beta, \epsilon_0, \epsilon_1)$-Gilbert-Elliot ($\GE$) packet erasure channel with a round-robin scheduling policy. We employ maximum distance separable ($\MDS$) scheme at the aggregator for encoding the multi-source updates. We characterize the mean $\AoI$ for the $\MDS$ coded system. Further, for large blocklengths, we show that the \emph{optimal coding rate} that achieves maximum \emph{coding gain} over the uncoded system is $1-\pers-\smallO(1)$, where $\pers \triangleq \frac{\beta}{\alpha+\beta}\epsilon_0 + \frac{\alpha}{\alpha+\beta}\epsilon_1$, and this maximum coding gain is $1+\pers -\smallO(1)$. 
\end{abstract}\vspace{-.6cm}
 \blfootnote{ S. Singhvi and P. D. Mankar  are with Signal Processing and Communication Research Center, IIIT Hyderabad, India. Email: shubhranshsinghvi2001@gmail.com,   praful.mankar@iiit.ac.in. }
\section{Introduction}
%In this work, we consider a multi-source status-update system connected to a destination node via a common aggregator node. The aggregator node sends coded status-update through an unreliable $\GE$ packet erasure channel. We examine how to employ coding redundancy in order to minimize an $\AoI$ metric. We compare the performance of two $\FEC$ techniques: $\MDS$ codes and \textcolor{red}{streaming codes} over the uncoded setting.
The sixth generation cellular network is envisioned to facilitate a massive machine-type communication (MMTC) \cite{Jiang} where a large number of sources send information/status updates of some physical random processes to  their destinations. For various use cases of MMTC, it is often crucial to ensure the {\em freshness or timely delivery} of status updates  at the destination. {\em Age of information} ($\AoI$) is  a key performance indicator  for measuring the freshness of updates received at the destination.
A vast amount of research work exists on the analytical characterization of $\AoI$ for a variety of systems including single source, multi-source, energy harvesting, multi-hop, distributed caching, distributed storage, etc. The interested reader may refer to \cite{RoyYates_2021Survey} for a detailed survey on the  analyses of $\AoI$.  
Most of these works  assume the perfect update delivery by ignoring the unreliability of communication channel. The $\AoI$ in the presence of transmission error essentially depends on the successful update delivery rate.
However, one can introduce  redundancy in the transmission to recover a few symbols/updates from the error and ensure successful delivery in a timely manner, which in turn can improve the $\AoI$ performance. Thus, it is important to investigate the benefits of error correction codes for controlling $\AoI$ over erasure channel.
Basically, the long codewords/blocks help to combat transmission errors but they increase the transmission time. However, the short ones require small transmission time but  has lower ability to correct the errors. As a result, there is a natural tension between the coding rate and $\AoI$ performance.
Inspired by this, we focus on exploring the impact of coding on the $\AoI$ performance in a multi-source update system.  

{\em Related Works}: The works studying the impact/benefits of error correction codes on  $\AoI$ are relatively spare in the literature. In \cite{ChenKun_2016},  $\AoI$ is investigated in the presence of error,   where the  update received after random service time is considered to be in error with some probability. The optimal transmission  of rateless codes over the erasure channel for minimizing $\AoI$ is characterized in \cite{FengSongtao_2019RatelessCode_OptimalAoI}.
The authors of \cite{RoyYates_2017} derived the mean $\AoI$  for fixed redundancy (FR) and infinite incremental redundancy (IIR) coded update transmission over a binary erasure channel (BEC). 
The authors have shown that FR codes, not requiring delivery feedback, can match the mean $\AoI$ performance of the IIR codes that does require feedback.  
Further, \cite{Xie_2020ARQ} studied the average $\AoI$ for FR low-density parity-check (LDPC) coded status update with and without automatic repeat request (ARQ). 
The interplay between the timeliness and the codeword length over BEC was investigated in \cite{ParagParimal_2017BEC}  for single transmission  and hybrid ARQ (HARQ) schemes. Here, in HARQ scheme, the update is encoded into $aN$ symbols of which subsequent $N$ symbols are transmitted in each slot until the update is decoded successfully. 
The authors of \cite{ElieNajm_2017HARQ} analyzed  $\AoI$ (with and without update preemption)   over BEC while employing FR-HARQ  and infinite incremental redundancy (IIR)-HARQ schemes for error correction. 
For a general distributed service time, the authors have shown that non preemption is a best policy for both FR-HARQ and IIR-HARQ schemes. 
 However, \cite{ElieNajm_2019OptimalAge} presented an information theoretic characterization of optimal achievable age over the erasure channel without delivery feedback.
The above works, except \cite{ElieNajm_2019OptimalAge},  are focused on characterizing the $\AoI$ specific to a coding scheme  for {\em a single source system}. 
On the other hand,  \cite{FengSongtao_2019Conding_BroadcastNetwork} investigated  the impact of coding on $\AoI$ in a broadcast system including two users and a single source with perfect feedback channels. Therein, the authors focused on designing an adaptive coding scheme to minimize the $\AoI$ at both the users. 
In \cite{Xingran_2019Conding_BroadcastNetwork}, the authors considered a system where the sender broadcasts coded updates from two streams to two users over erasure channel with feedback. 
It was shown that the coded randomized policies   provide improved $\AoI$ performance compared to the corresponding uncoded policies. Further, in \cite{FaraziShahab_2020}, mean $\AoI$ is derived  for multi-source update system with stationary randomized and round-robin scheduling  with various retransmission policies. All above works, except \cite{ElieNajm_2019OptimalAge},  assumed   {\em a perfect delivery feedback channel} from the destination to the source/sender, which may not be practical for system including MMTC. From the practical  implementation perspectives, it may be desirable to design an age optimal coding scheme for a multi-source update system without feedback channel, which is the main goal of this paper. 

{\em Contributions:} 
This paper considers  a multi-source update system where the  updates from the multiple sources are transmitted to a common destination via some aggregator, as shown in Fig. \ref{fig:sys_mod}. We model the packet erasure channel as the commonly-accepted $(\alpha, \beta, \epsilon_0, \epsilon_1)-\GE$ channel \cite{gilbert}-\cite{HasHoh} and assume no feedback from the destination  to the aggregator or source. We employ $\MDS$ codes at the aggregator for encoding the multi-source updates. For this setting, we first derive the mean $\AoI$ and next we analytically characterize the gain in mean $\AoI$  that is achievable by $\MDS$ coded system over uncoded system for large blocklengths. In addition, we also study the interplay between the coding rate and $\AoI$ gain, and derive the optimal coding rate that maximizes the $\AoI$ gain. In particular, we show that the optimal coding rate that achieves maximum coding gain over the uncoded system is $1-\pers-\smallO(1)$, where $\pers \triangleq \frac{\beta}{\alpha+\beta}\epsilon_0 + \frac{\alpha}{\alpha+\beta}\epsilon_1$, and this maximum coding gain is  $1+\pers-\smallO(1)$. 
\vspace{-0.15cm}
\section{System Model}
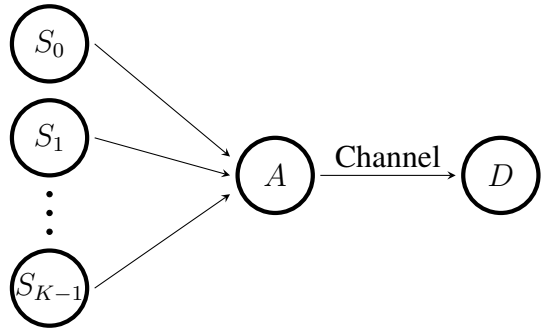
\begin{figure}[H]
\centering
\resizebox{4cm}{!}
{
\begin{tikzpicture}
\begin{large}
\draw [ultra thick](0,-0.25) circle (0.5);
\draw (0,-0.25) node{$S_0$};
\draw [ultra thick](0,-1.5) circle (0.5);
\draw (0,-1.5) node{$S_1$};
\filldraw (0,-2.25) circle (1pt);
\filldraw (0,-2.5) circle (1pt);
\filldraw (0,-2.75) circle (1pt);
\draw [ultra thick](0,-3.5) circle (0.5);
\draw (0,-3.5) node{$S_{K-1}$};
\draw [-stealth] (0.6,-0.25) -- (2.4,-1.75);
\draw [-stealth] (0.6,-1.5) -- (2.4,-2);
\draw [-stealth] (0.6,-3.5) -- (2.4,-2.25);

\draw [ultra thick](3,-2) circle (0.5);
\draw (3,-2) node{$A$};

\draw [-stealth] (3.6,-2) -- (5.4,-2);
\draw (4.5,-1.75) node{$\text{Channel}$};

\draw [ultra thick](6,-2) circle (0.5);
\draw (6,-2) node{$D$};
\end{large}
\end{tikzpicture}
}
\caption{Multi-source Update Communication System.}
\label{fig:sys_mod}
\end{figure}

We consider a status-update system with $K$ sources $S_i$ where $i\in [0:K-1]$, an aggregator node $A$ and a destination node $D$ as shown in Fig~\ref{fig:sys_mod}. The source nodes wish to communicate time-sensitive status updates of their associated random processes with the destination by transmitting packets of size $\ell$ symbols. 

We consider that each source can generate the packets \textit{at will} and communicate them to the aggregator within a negligible time. 
The aggregator is assumed to transmit the stream of packets to the destination in a slotted manner such that one time slot per symbol is required. 
%We assume that the time of transmission between the source and aggregator is negligible and that Time is divided into slots of equal duration, such that it takes a time slot to transmit one symbol. 
We also assume that the entire packet is generated before the first symbol of the packet is transmitted. Further, the aggregator is assumed to apply a round-robin scheduling  for forwarding the packets from $K$ sources to the destination. %, thus the source nodes are selected deterministically in order. 
Let $\{\B{m}_i\}_{i=0}^{\infty}$ denote the stream of packets received by the aggregator, where $\B{m}_i\in\mathcal{M}$ and $\mathcal{M}$ is the alphabet of $\ell$ symbols.
%The aggregator node receives a stream of samples , where $\B{m}_i\in{\{0,1,\ldots,q-1\}}^\ell$ is transmitted by $S_{\{i\mod K\}}$. 
%For $i\in[0:K-1]$, let $\delta_i$ denote the instantaneous $\AoI$ of the source $S_i$. We next define $\delta_i$ as 

The  $\AoI$ of the source $S_i$ at time instance $t$ is defined as
\bea
\label{inst_AoI}
\delta_i(t)  \triangleq t - u_i(t),
\eea
where $u_i(t)$ denotes the time stamp of the generation of the latest packet of the source $S_i$. Thus, the  average $\AoI$ for the source $S_i$ can be evaluated as
\bea
\label{avg_AoI}
\Delta_i = \underset{\tau\to\infty}{\lim} \frac{1}{\tau} \int_0^{\tau} \delta_i(t) dt.
\eea
We now describe the uncoded and the coded transmission policies adopted by the aggregator node in the following subsections. We will refer to the former as the {\em uncoded status-update system} and the latter as the {\em coded status-update system}. 
\subsection{Uncoded Status-update System}
In the uncoded update system, the aggregator simply forwards the stream of packets from $K$ sources in a round-robin fashion. If a packet from source $S_i$ is not successfully received at the destination, then it is discarded and a new packet is transmitted during the next scheduled slot for the source $S_i$. Therefore, the $\AoI$ becomes equal to $\ell$ units of transmission slot duration after every successful transmission in this system.  
\subsection{Coded Status-Update Systems}
\begin{figure*}
\centering
\resizebox{.65\textwidth}{!}
{
\begin{tikzpicture}
\draw [ultra thick, draw=black, fill=red, opacity=0.2] (-2.35,4) rectangle (-0.35,4.5); 
\draw [ultra thick, draw=black, fill=blue, opacity=0.2] (-2.35,3.5) rectangle (-0.35,4); 
\draw [ultra thick, draw=black, fill=green, opacity=0.2] (-2.35,3) rectangle (-0.35,3.5); 
\draw [opacity=0.2] (-2.35,2.5) rectangle (-0.35,3); 
\draw [ultra thick, draw=black, fill=orange, opacity=0.2] (-2.35,2) rectangle (-0.35,2.5); 
\draw (-1.35,4.25) node{\scriptsize{$\B{m}_{i}(0)$}};
\draw (-1.35,3.75) node{\scriptsize{$\B{m}_{i}(1)$}};
\draw (-1.35,3.25) node{\scriptsize{$\B{m}_{i}(2)$}};
\draw (-1.35,2.9) node{\scriptsize{$.$}};
\draw (-1.35,2.75) node{\scriptsize{$.$}};
\draw (-1.35,2.6) node{\scriptsize{$.$}};
\draw (-1.35,2.25) node{\scriptsize{$\B{m}_{i}(\ell-1)$}};

\draw [ultra thick, draw=black, fill=red, opacity=0.2] (-0.35,4) rectangle (1.65,4.5); 
\draw [ultra thick, draw=black, fill=blue, opacity=0.2] (-0.35,3.5) rectangle (1.65,4); 
\draw [ultra thick, draw=black, fill=green, opacity=0.2] (-0.35,3) rectangle (1.65,3.5); 
\draw [opacity=0.2] (-0.35,2.5) rectangle (1.65,3); 
\draw [ultra thick, draw=black, fill=orange, opacity=0.2] (-0.35,2) rectangle (1.65,2.5); 
\draw (0.65,4.25) node{\scriptsize{$\B{m}_{i+1}(0)$}};
\draw (0.65,3.75) node{\scriptsize{$\B{m}_{i+1}(1)$}};
\draw (0.65,3.25) node{\scriptsize{$\B{m}_{i+1}(2)$}};
\draw (0.65,2.9) node{\scriptsize{$.$}};
\draw (0.65,2.75) node{\scriptsize{$.$}};
\draw (0.65,2.6) node{\scriptsize{$.$}};
\draw (0.65,2.25) node{\scriptsize{$\B{m}_{i+1}(\ell-1)$}};

\filldraw (2.3,3.25) circle (1pt);
\filldraw (2.65,3.25) circle (1pt);
\filldraw (3,3.25) circle (1pt);

\draw [ultra thick, draw=black, fill=red, opacity=0.2] (3.65,4) rectangle (5.65,4.5); 
\draw [ultra thick, draw=black, fill=blue, opacity=0.2] (3.65,3.5) rectangle (5.65,4); 
\draw [ultra thick, draw=black, fill=green, opacity=0.2] (3.65,3) rectangle (5.65,3.5); 
\draw [opacity=0.2] (3.65,2.5) rectangle (5.65,3); 
\draw [ultra thick, draw=black, fill=orange, opacity=0.2] (3.65,2) rectangle (5.65,2.5); 
\draw (4.65,4.25) node{\scriptsize{$\B{m}_{i+k-1}(0)$}};
\draw (4.65,3.75) node{\scriptsize{$\B{m}_{i+k-1}(1)$}};
\draw (4.65,3.25) node{\scriptsize{$\B{m}_{i+k-1}(2)$}};
\draw (4.65,2.9) node{\scriptsize{$.$}};
\draw (4.65,2.75) node{\scriptsize{$.$}};
\draw (4.65,2.6) node{\scriptsize{$.$}};
\draw (4.65,2.25) node{\scriptsize{$\B{m}_{i+k-1}(\ell-1)$}};

\draw [ultra thick, draw=black, fill=red, opacity=0.2] (5.65,4) rectangle (7.65,4.5); 
\draw [ultra thick, draw=black, fill=blue, opacity=0.2] (5.65,3.5) rectangle (7.65,4); 
\draw [ultra thick, draw=black, fill=green, opacity=0.2] (5.65,3) rectangle (7.65,3.5); 
\draw [opacity=0.2] (5.65,2.5) rectangle (7.65,3); 
\draw [ultra thick, draw=black, fill=orange, opacity=0.2] (5.65,2) rectangle (7.65,2.5); 
\draw (6.65,4.25) node{\scriptsize{$\B{p}_{1}(0)$}};
\draw (6.65,3.75) node{\scriptsize{$\B{p}_{1}(1)$}};
\draw (6.65,3.25) node{\scriptsize{$\B{p}_{1}(2)$}};
\draw (6.65,2.9) node{\scriptsize{$.$}};
\draw (6.65,2.75) node{\scriptsize{$.$}};
\draw (6.65,2.6) node{\scriptsize{$.$}};
\draw (6.65,2.25) node{\scriptsize{$\B{p}_{1}(\ell-1)$}};

\filldraw (8.3,3.25) circle (1pt);
\filldraw (8.65,3.25) circle (1pt);
\filldraw (9,3.25) circle (1pt);

\draw [ultra thick, draw=black, fill=red, opacity=0.2] (9.65,4) rectangle (11.65,4.5); 
\draw [ultra thick, draw=black, fill=blue, opacity=0.2] (9.65,3.5) rectangle (11.65,4); 
\draw [ultra thick, draw=black, fill=green, opacity=0.2] (9.65,3) rectangle (11.65,3.5); 
\draw [opacity=0.2] (9.65,2.5) rectangle (11.65,3); 
\draw [ultra thick, draw=black, fill=orange, opacity=0.2] (9.65,2) rectangle (11.65,2.5); 
\draw (10.65,4.25) node{\scriptsize{$\B{p}_{n-k}(0)$}};
\draw (10.65,3.75) node{\scriptsize{$\B{p}_{n-k}(1)$}};
\draw (10.65,3.25) node{\scriptsize{$\B{p}_{n-k}(2)$}};
\draw (10.65,2.9) node{\scriptsize{$.$}};
\draw (10.65,2.75) node{\scriptsize{$.$}};
\draw (10.65,2.6) node{\scriptsize{$.$}};
\draw (10.65,2.25) node{\scriptsize{$\B{p}_{n-k}(\ell-1)$}};

\end{tikzpicture}
}
\caption{Illustrating a coded block at the aggregator node. Here, $(\cdot)$ denotes the symbol index in a packet.}\vspace{-3mm}
\label{fig:cod_stream}
\end{figure*}
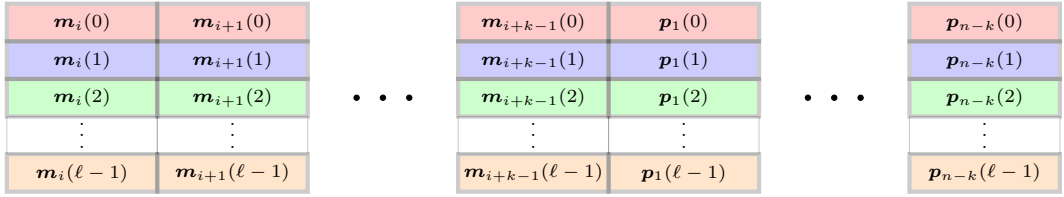

In the coded update system, the aggregator node is assumed to employ an $(n,k)-\MDS$ code across the stream of packets. For the ease of analysis, we consider $K=\lambda k$, where $\lambda\in\mathbb{Z}_+$, i.e., the stream of packets at the aggregator is divided into $\lambda$ \textit{blocks} each containing $k$ pakects. In each block, the idea is to horizontally embed $\ell$ codewords of a systematic $(n,k)-\MDS$ to form the \textit{coded block} of $n$ packets each containing $\ell$ symbols as shown in the Fig.~\ref{fig:cod_stream}. The last $n-k$ packets will be referred to as \textit{parity} packets. Due to the $\MDS$ property, a coded block can tolerate $n-k$ packet erasures. 
%When a coded block suffers less than $n-k$ packet erasures, for the ease of analysis, we assume that if a sample packet is lost then it can be recovered only after the complete block is transmitted. 
Thus, if a packet transmission is in error, then it can be recovered from the coded block only when the block suffers less than or equal to $n-k$ packets erasures. In such cases, we assume that the packet is recovered at the end of block for the ease of analysis.
In the event of a block failure (i.e.,  more than $n-k$ erasures occur in a block), the aggregator discards the failed/lost packet and uses new packet for transmitting at the corresponding position in the next scheduled block. %then the lost samples are discarded. 
%In the next section, we describe this Block Error Probability ($\BEP$) of an $(n,k)-\MDS$ code for the $\GE$ packet erasure channel. 
For evaluating the average $\AoI$, analytical characterization of the block error probability ($\BEP$) of the above discussed  $(n,k)-\MDS$ code for a given class of erasure channel is necessary. In the next section, we describe the channel model and also examine  $\BEP$  for it.

\section{Channel Model and Block Error Probability}
\label{CMBEP}
\begin{figure}[H]
    \centering
\includegraphics[width=0.32\textwidth]{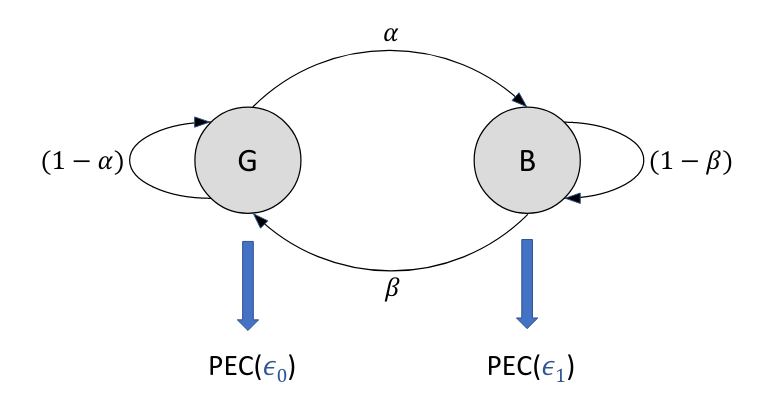}
    \caption{$(\alpha, \beta, \epsilon_0, \epsilon_1)$-Gilbert-Elliot Channel Model}
    \label{fig:GE}
\end{figure}

The channel between the aggregator node and the destination node is modeled as a $\GE$  erasure channel which is a two-state Markov channel characterized by the parameters $(\alpha,\beta, \epsilon_0, \epsilon_1)$. %The channel has two states, good and bad. 
The $\GE$ channel toggles between two states, good and bad.
In the good state denoted by $\mathsf{G}$, the channel is $\PEC(\epsilon_0)$ and in the bad state denoted by $\mathsf{B}$, the channel is $\PEC(\epsilon_1)$, where  $\PEC(\epsilon)$ is  a  packet erasure  channel  with  packet  erasure  probability $\epsilon$, and $\alpha$ and  $\beta$ are the state transition  probabilities as  shown in the Fig. \ref{fig:GE}. Interested readers can find some other works pertaining to the design of forward error-correcting codes for various models of communication over the $\GE$ channel in \cite{MartTrotISIT07}-\cite{HannaAnt23}.

\noindent Let $P(n,e)$ denote the probability of seeing $e$ erasures in $n$ instances of the $\GE$ channel. The closed form expression for $P(n,e)$ is presented in \cite{VajRamJhaKum20,VajRamJhaKum21}, but for completeness, we provide its detailed analytical characterization. Let $\mathsf{S}_t\in \{0,1\}$ and $\mathsf{E}_t \in \{0,1\}$ denote the state of GE channel and  the presence of erasure at time $t$, respectively, where $\mathsf{E}_t = 1$ indicates an erasure at time $t$, and $\mathsf{S}_t = 1$ indicates the bad state at time $t$. Thus, we have $P(\mathsf{E}_t = 1 | \mathsf{S}_t = 0) = \epsilon_0$ and $P(\mathsf{E}_t = 1 | \mathsf{S}_t = 1) = \epsilon_1$. Let $\pi_g \triangleq \frac{\beta}{\alpha+\beta}$ denote the steady state probability of good state. Let the probability of seeing $e$ erasures in an $n$ length window $[t+1, t+n]$ when $(t+n)$-th state is $0$ and $1$ be given by $g(n,e, \pi_g)$ and $b(n,e, \pi_g)$, respectively. % \textcolor{red}{Let $g(n,e,\pi_g) \triangleq P(w(\mathsf{E}_{t+1}^{t+n}) = e, \mathsf{S}_{t+n} = 0)$ and $b(n,e,\pi_g) \triangleq P(w(\mathsf{E}_{t+1}^{t+n}) = e, \mathsf{S}_{t+n} = 1)$.} \textcolor{blue}{Some description of functions $g$  and $b$ can be useful.} 
% This is clearly independent of $t$ given steady state probability $\pi_g$. Thus, we have 
%  \bea
%  \label{eq:pnk}
%  P(n,e) = g(n,e,\pi_g)+b(n,e,\pi_g).
%  \eea
The probabilities $g$ and $b$ can be computed as 
\begin{align}
\label{eq:gbnk}
\begin{split}
\left[\begin{array}{c}
    g(n,e, \pi_g)\\
    b(n,e, \pi_g)
\end{array}\right] &= \sum\limits_{r=0}^n \sum\limits_{b=(e+r-n)^+}^{\min\{r, e\}} {r \choose b}{n-r \choose e-b} (\bar{\epsilon}_0)^{n+b-e-r} \cdot\\
&\epsilon_0^{e-b}(\bar{\epsilon}_1)^{r-b}\epsilon_1^b \left[\begin{array}{c}
	g_B(n,r, \pi_g)\\
	b_B(n,r, \pi_g)
\end{array}\right],
\end{split}
\end{align}
% \begin{figure*}[!h]
% \begin{align}
% \begin{split}
% \label{eq:gbnk}
% \left[\begin{array}{c}
%     g(n,e, \pi_g)\\
%     b(n,e, \pi_g)
% \end{array}\right] = 
% \sum\limits_{r=0}^n \sum\limits_{b=(e+r-n)^+}^{\min\{r, e\}} {r \choose b}{n-r \choose e-b} (\bar{\epsilon}_0)^{n+b-e-r} \epsilon_0^{e-b}(\bar{\epsilon}_1)^{r-b}\epsilon_1^b \left[\begin{array}{c}
% 	g_B(n,r, \pi_g)\\
% 	b_B(n,r, \pi_g)
% \end{array}\right].
% \end{split}
% \end{align}
% \end{figure*}
where $g_B(n,r,\pi_g)$  and $b_B(n,r,\pi_g)$ are the probabilities  of  being  in  bad  state  for $r$ out  of $n$ instances and  ending  in  good  and bad  states, respectively.  % $g_B(n,r,\pi_g)= P(w(\mathsf{S}_{t+1}^{t+n}) = r, \mathsf{S}_{t+n} = 0)$ and $b_B(n,r,\pi_g)= P(w(\mathsf{S}_{t+1}^{t+n}) = r, \mathsf{S}_{t+n} = 1)$. 
These probabilities  can be derived as
\begin{small}
\begin{align}
\label{eq:gbfromm} \left[\begin{array}{c}
g_B(n,r, \pi_g)\\
b_B(n,r, \pi_g)
\end{array}\right] = \left[\begin{array}{ccc}
\pi_g & \beta\bar{\pi_g} & -\bar{\beta} \pi_g\\
\bar{\pi_g} & -\bar{\alpha}\bar{\pi_g} & \alpha \pi_g\\
\end{array}\right] \left[\begin{array}{c}
m(n,r)\\
m(n-1,r)\\
m(n-1,r-1)
\end{array}\right],
\end{align}
\end{small}
\vspace{-0.6cm}

\noindent where $m(n,r) =$
\begin{align}
\begin{split}
\label{eq:mnr}   \sum\limits_{a = \max\{r,n-r\}}^n {a \choose r} {r \choose n-a} (\alpha-\bar{\beta})^{n-a} \bar{\beta}^{a-n+r} \bar{\alpha}^{a-r}.
\end{split}
\end{align}
Note that $\alpha \geq 1-\beta$ is needed to ensure \eqref{eq:mnr} is positive. If it's not the case, then we can simply reverse the states.
% \begin{lemma}
%     From \eqref{eq:pnk}, \eqref{eq:gbnk}, \eqref{eq:gbfromm} and \eqref{eq:mnr} we have an analytical expression for computation of $P(n,e)$.
% \end{lemma}
\begin{lemma}
    The probability of observing $e$ erasures in $n$ uses of the $\GE$ channel is given by
    \bea
 \label{eq:pnk}
 P(n,e) = g(n,e,\pi_g)+b(n,e,\pi_g).
 \eea
where the probabilities $g$ and $b$ can be calculated using \eqref{eq:gbnk}-\eqref{eq:mnr}.
\end{lemma}
% \textcolor{blue}{If possible, try to remove (3). Perhaps, including one or two sentences of functions $g$ and $b$, as suggested above, can allow to remove (3).}

By substituting $n=1$ and $e=1$ in the expression for $P(n,e)$, the next corollary follows. 
\begin{corollary}
\label{cor:pe}
The probability of observing one erasure over the $(\alpha,\beta, \epsilon_0, \epsilon_1)-\GE$ channel is
\bea
\label{pe}
\pers \triangleq P(1,1)=\epsilon_0\pi_g + \epsilon_1\bar{\pi}_g.
\eea
\end{corollary}

\begin{corollary}
\label{cor:BEP}
The $\BEP$ for $(n,k)-\MDS$ code over the $(\alpha,\beta, \epsilon_0, \epsilon_1)-\GE$ channel is
\bea
\label{bep}
\BEP_{\MDS}(n,k) = 1 - \sum_{e=0}^{n-k} P(n,e).
\eea    
\begin{proof}
The result follows from the fact that an $(n,k)-\MDS$ code can not recover from any erasure pattern that lies outside the following set
\begin{align*}
\centering
\left\{\left(\mathsf{E}_t, \mathsf{E}_{t+1}, \ldots, \mathsf{E}_{t+n-1}\right): \sum_{i =t}^{t+n-1}\mathsf{E}_i\geq n-k\right\},     
\end{align*}
where the packets from $[t:t+n-1]$ are part of the same coded block.  
\end{proof}
\end{corollary}
% \begin{corollary}
% \label{albet1}
% For $\alpha = 1-\beta + \smallO(1)$, 
% \bean
% P(n,e) = \binom{n}{e}\pers^e(1-\pers)^{n-e}.
% \eean
% \end{corollary}
% \begin{proof}
% For $\alpha = 1-\beta + \smallO(1)$, $\pi_g = \beta$ and from \eqref{eq:mnr}, $m(n,r) \Tilde{=} \binom{n}{r}\alpha^r\beta^{n-r}.$ From \eqref{eq:gbfromm}, it can be verified that $g_B(n,r,\pi_g) + b_B(n,r,\pi_g) \Tilde{=} m(n,r)$. Therefore, from $\eqref{eq:gbnk}$, we have that $P(n,e) =\sum\limits_{r=0}^n\sum\limits_{b=(e+r-n)^+}^{\min\{r, e\}} {r \choose b}{n-r \choose e-b} (\bar{\epsilon}_0)^{n+b-e-r} \cdot \epsilon_0^{e-b}(\bar{\epsilon}_1)^{r-b}\epsilon_1^b .m(n,r)$. 
% \end{proof}
\section{Analysis of Average $\AoI$}
\begin{figure}[H]
\centering
\resizebox{.4 \textwidth}{!}
{
\begin{tikzpicture}
\begin{footnotesize}
\draw [-stealth](0,0) -- (0,4);
\draw [-stealth](0,0) -- (8,0);
\draw (-0.05,0.15) -- (0.05,0.15);

\draw (-0.25,0.15) node{$\ell$};
\draw (-0.05,1.65) -- (0.05,1.65);
\draw (-0.8,1.65) node{$(K+1)\ell$};
\draw (-0.05,3.15) -- (0.05,3.15);
\draw (-0.825,3.15) node{$(2K+1)\ell$};

\draw (-0.05, 0.15) -- (0.05, 0.15);
\draw (-0.05, 0.30) -- (0.05, 0.30);
\draw (-0.05,0.45) -- (0.05,0.45);
\draw (-0.05,0.60) -- (0.05,0.60);
\draw (-0.05,0.75) -- (0.05,0.75);
\draw (-0.05,0.90) -- (0.05,0.90);
\draw (-0.05,1.05) -- (0.05,1.05);
\draw (-0.05,1.20) -- (0.05,1.20);
\draw (-0.05,1.35) -- (0.05,1.35);
\draw (-0.05,1.50) -- (0.05,1.50);
\draw (-0.05,1.65) -- (0.05,1.65);
\draw (-0.05,1.80) -- (0.05,1.80);
\draw (-0.05,1.95) -- (0.05,1.95);
\draw (-0.05,2.10) -- (0.05,2.10);
\draw (-0.05,2.25) -- (0.05,2.25);
\draw (-0.05,2.40) -- (0.05,2.40);
\draw (-0.05,2.55) -- (0.05,2.55);
\draw (-0.05,2.70) -- (0.05,2.70);
\draw (-0.05,2.85) -- (0.05,2.85);
\draw (-0.05,3.00) -- (0.05,3.00);
\draw (-0.05,3.15) -- (0.05, 3.15);
\draw (-0.05,3.30) -- (0.05, 3.30);
\draw (-0.05,3.45) -- (0.05, 3.45);
\draw (-0.05,3.60) -- (0.05, 3.60);
\draw (-0.05,3.75) -- (0.05, 3.75);

\draw (0.15,-0.25) node{$\ell$};
\draw (1.65,-0.05) -- (1.65,0.05);
\draw (1.65,-0.25) node{$(K+1)\ell$};
\draw (3.15,-0.05) -- (3.15,0.05);
\draw (3.15,-0.25) node{$(2K+1)\ell$};
\draw (4.65,-0.05) -- (4.65,0.05);
\draw (4.65,-0.25) node{$(3K+1)\ell$};
\draw (6.15,-0.05) -- (6.15,0.05);
\draw (6.15,-0.25) node{$(4K+1)\ell$};
\draw (7.65,-0.05) -- (7.65,0.05);
\draw (7.65,-0.25) node{$(5K+1)\ell$};

\draw (0.15,-0.05) -- (0.15,0.05);
\draw (0.30,-0.05) -- (0.30,0.05);
\draw (0.45,-0.05) -- (0.45,0.05);
\draw (0.60,-0.05) -- (0.60,0.05);
\draw (0.75,-0.05) -- (0.75,0.05);
\draw (0.90,-0.05) -- (0.90,0.05);
\draw (1.05,-0.05) -- (1.05,0.05);
\draw (1.20,-0.05) -- (1.20,0.05);
\draw (1.35,-0.05) -- (1.35,0.05);
\draw (1.50,-0.05) -- (1.50,0.05);
\draw (1.65,-0.05) -- (1.65,0.05);
\draw (1.80,-0.05) -- (1.80,0.05);
\draw (1.95,-0.05) -- (1.95,0.05);
\draw (2.10,-0.05) -- (2.10,0.05);
\draw (2.25,-0.05) -- (2.25,0.05);
\draw (2.40,-0.05) -- (2.40,0.05);
\draw (2.55,-0.05) -- (2.55,0.05);
\draw (2.70,-0.05) -- (2.70,0.05);
\draw (2.85,-0.05) -- (2.85,0.05);
\draw (3.00,-0.05) -- (3.00,0.05);
\draw (3.15,-0.05) -- (3.15,0.05);
\draw (3.30,-0.05) -- (3.30,0.05);
\draw (3.45,-0.05) -- (3.45,0.05);
\draw (3.60,-0.05) -- (3.60,0.05);
\draw (3.75,-0.05) -- (3.75,0.05);
\draw (3.90,-0.05) -- (3.90,0.05);
\draw (4.05,-0.05) -- (4.05,0.05);
\draw (4.20,-0.05) -- (4.20,0.05);
\draw (4.35,-0.05) -- (4.35,0.05);
\draw (4.50,-0.05) -- (4.50,0.05);
\draw (4.65,-0.05) -- (4.65,0.05);
\draw (4.80,-0.05) -- (4.80,0.05);
\draw (4.95,-0.05) -- (4.95,0.05);
\draw (5.10,-0.05) -- (5.10,0.05);
\draw (5.25,-0.05) -- (5.25,0.05);
\draw (5.40,-0.05) -- (5.40,0.05);
\draw (5.55,-0.05) -- (5.55,0.05);
\draw (5.70,-0.05) -- (5.70,0.05);
\draw (5.85,-0.05) -- (5.85,0.05);
\draw (6.00,-0.05) -- (6.00,0.05);
\draw (6.15,-0.05) -- (6.15,0.05);
\draw (6.30,-0.05) -- (6.30,0.05);
\draw (6.45,-0.05) -- (6.45,0.05);
\draw (6.60,-0.05) -- (6.60,0.05);
\draw (6.75,-0.05) -- (6.75,0.05);
\draw (6.90,-0.05) -- (6.90,0.05);
\draw (7.05,-0.05) -- (7.05,0.05);
\draw (7.20,-0.05) -- (7.20,0.05);
\draw (7.35,-0.05) -- (7.35,0.05);
\draw (7.50,-0.05) -- (7.50,0.05);
\draw (7.65,-0.05) -- (7.65,0.05);
\draw (7.80,-0.05) -- (7.80,0.05);

\draw (0,-0.75) node{$T_0$};

\draw [ultra thick, black] (0,0) -- (1.65,1.65);
\draw [ultra thick, black] (1.65,1.65) -- (1.65,0.15); 
\draw [|-|](0.2,-0.75) -- (1.45,-0.75);
\draw (1.65,-0.75) node{$T_1$};
\draw (0.825,-1) node{$X_1$};
\draw (1,0.5) node{$A_1$};

\draw [ultra thick, black] (1.65,0.15) -- (4.65,3);
\draw [ultra thick, black] (4.65,3) -- (4.65,0.15); 
\draw [|-|](1.85,-0.75) -- (4.45,-0.75);
\draw (4.65,-0.75) node{$T_2$};
\draw (3.15,-1) node{$X_2$};
\draw (4,2) node{$A_2$};

\draw [ultra thick, black] (4.65,0.15) -- (6.15,1.65);
\draw [ultra thick, black] (6.15,1.65) -- (6.15,0.15); 
\draw [|-|](4.85,-0.75) -- (5.95,-0.75);
\draw (6.15,-0.75) node{$T_3$};
\draw (5.4,-1) node{$X_3$};
\draw (5.5,0.5) node{$A_3$};

\draw [-stealth, opacity = 0.5, red] (0,-0.1) -- (0,-0.5);
\draw [-stealth, opacity = 0.5, red ] (1.65,-0.1) -- (1.65,-0.5);
\draw [-stealth, opacity = 0.5, red] (4.65,-0.1) -- (4.65,-0.5);
\draw [-stealth, opacity = 0.5, red] (6.15,-0.1) -- (6.15,-0.5);

\draw [ultra thick, black] (6.15,0.15) -- (7.65,1.65);
\end{footnotesize}
\draw (0,4.25) node{$\delta_1(t)$};
\draw (8.25,0) node{$t$};
\end{tikzpicture}
}
\caption{Illustrating a realisation of the instantaneous $\AoI$ $(\delta_0)$ for the uncoded system.}
\label{fig:AoI-uncoded}
\end{figure}
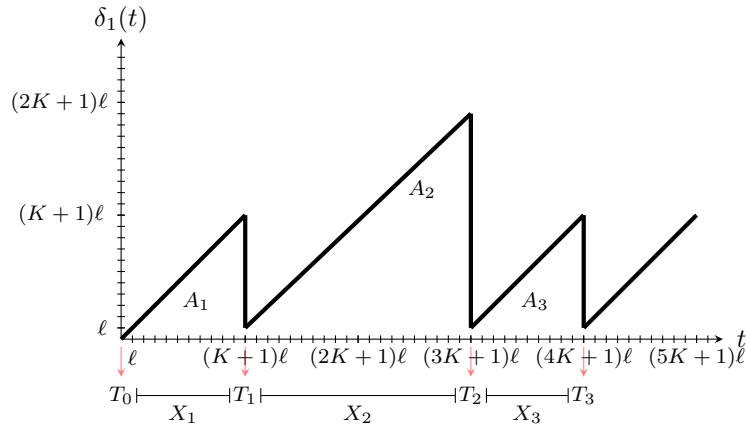

In this section, we derive the average $\AoI$ for a source $S_i$ for the uncoded and coded systems. For $j \in \mathbb{N}$, let the random variable $\mathsf{T}_j$ denote the time at the $j$-th successful reception of a packet from $S_i$. We set $\mathsf{T}_0 \triangleq i\ell$. For $j \in \mathbb{N}$, let the random variable $\mathsf{X}_j$ denote the inter-arrival time between the $(j-1)$-th and $j$-th successful receptions, i.e., $\mathsf{X}_j = \mathsf{T}_j - \mathsf{T}_{j-1}$. To analyze the average $\AoI$, we use the geometric method wherein we decompose the area defined by \eqref{avg_AoI} into a sum of disjoint polygons $A_1, A_2, \ldots$ as shown in Fig. \ref{fig:AoI-uncoded}. 
Using this decomposition, we derive $\Delta_i$ as follows \cite{RoyYates_2017}
\bea
\label{geo_AoI}
\Delta_i = \underset{\ell\to\infty}{\lim} \frac{1}{T_\ell} \sum_{j=1}^{\ell} A_j=\underset{\ell\to\infty}{\lim}\dfrac{\frac{1}{l}\sum_{j=1}^{\ell}A_j}{\frac{1}{l}\sum_{j=1}^{\ell}\mathsf{X}_j}= \dfrac{\mathbb{E}[A]}{\mathbb{E}[X]}. \eea
From Fig. \ref{fig:AoI-uncoded}, it can be seen that
\bea
A_j = \frac{(\mathsf{X}_j+\ell)^2}{2} - \frac{\ell^2}{2} = \frac{\mathsf{X}_j^2}{2} + \mathsf{X}_j\ell.
\eea
Since  $\mathsf{X}_j$s are independent and identically distributed (i.i.d.), $\mathbb{E}[A] = 0.5\mathbb{E}[X^2]+ \ell\mathbb{E}[X]$. Thus, using \eqref{geo_AoI}, we evalaute the average $\AoI$ for the source $S_i$ as
\begin{equation}
\Delta_i = \dfrac{\mathbb{E}[X^2]}{2\mathbb{E}[X]} + \ell.
\label{eq:AvgAoI}
\end{equation} 
\subsection{Analysis for Uncoded System}
In the uncoded status-update system, if a packet is not successfully received then it's discarded. The probability that  a packet is not successfully received is $\pers$, which is given in Corollary \ref{cor:pe}.
%Thus,  the random variable $\mathsf{X}_j$ is a geometric random variable with mean $\frac{K\ell}{1-\pers}$, 
The inter-arrival time of receptions is $X_j=xK\ell$ when $x$ is the number of  transmission attempts between successful transmissions for a generic source $S_i$. As $x$ follows geometric distribution, the probability mass function (pmf) of $\mathsf{X}_j$ becomes
\begin{align}
P(\mathsf{X}_j=xK\ell) = (1-\pers)\pers^{x-1}.
    \label{eq:pmf_X_uncoded}
\end{align}
If the packet is successfully received then the 
instantaneous $\AoI$ drops to $\ell$ as it takes $\ell$ slots to transmit a packet. Thus, the instantaneous $\AoI$ behaves like a sawtooth function as shown in Fig. \ref{fig:AoI-uncoded}. 
Using \eqref{eq:pmf_X_uncoded}, we can obtain $\mathbb{E}[X] = \frac{K\ell}{1-\pers}$ and $\text{Var}[X] = \frac{\pers K^2\ell^2}{(1-\pers)^2}$.  From this and \eqref{eq:AvgAoI}, our next lemma follows.
\begin{lemma}
\label{lemma:Age_uncoded}
The average $\AoI$ of source $S_i$ for the uncoded update system is
\bea
\Delta_i = \dfrac{K\ell(1+\pers)}{2(1-\pers)} + \ell.
\eea
If $K\gg\frac{2(1-\pers)}{1+\pers}$, the average $\AoI$ for a generic source  can be approximated as
$$\Delta_{\mathsf{Uncoded}} \approx \dfrac{K\ell(1+\pers)}{2(1-\pers)}.$$
\end{lemma}
% \textcolor{red}{See $p_0$ is defined or not}.
%Consider the $s$-th sample of the source $S_i$, it belongs to the $((s-1)\lambda + \floor{\frac{i}{k}})$-th block and is at $\{i \mod k\}$-th position of the block. 
\vspace{-0.5cm}
\subsection{Analysis for  $(n,k)-\MDS$ Coded System}
As per the code construction discussed in Section II, the $s$-th packet of source $S_i$ belongs to $((s-1)\lambda + \floor{\frac{i}{k}})$-th block and is located in $\{i \mod k\}$-th position of that block.
Let this block be denoted by $\{\B{c}_0, \B{c}_1, \ldots, \B{c}_{n-1}\}$, where $\B{c}_i = \B{m}_i$ for $i \in [0:k-1]$ and $\B{c}_i=p_{i-k+1}$ for $i \in [k:n-1]$. Let $\mathsf{E}_t\in\{0,1\}$ for $0 \leq t \leq n-1$ indicate whether the coded packet $\B{c}_t$ is erased or not, here $1$ indicates an erasure. Now, consider the  following three events: 
\begin{itemize}
    \item Event $\cA=\{\mathsf{E}_{\{i \mod k\}} = 0\}$: {\em No erasure  and thus the age $\delta_i$ given in (1) drops to  $\ell$ as it takes $\ell$ slots to transmit a packet. From \eqref{pe}, we get} 
\begin{align}
\label{P_EA}
P(\cA) = 1-\pers.    
\end{align}
\item Event $\cB=\{\mathsf{E}_{\{i \mod k\}} = 1, \sum_{t=1}^n \mathsf{E}_t \leq n-k\}$: {\em The lost packet can be recovered from the block as the number of erasures are less than or equal to $n-k$. Since the packet is recovered at the end of block, the age $\delta_i$ drops to $(n+1-\{i \mod k\})\ell$. Using Corollary \ref{cor:pe} and \ref{cor:BEP}, we can write }
\begin{align}
\label{P_EB}
P(\cB) = \pers\left(1-\BEP_\MDS(n-1,k)\right).    
\end{align}
\item Event $\cC=\{\mathsf{E}_{\{i \mod k\}} = 1, \sum_{t=1}^n \mathsf{E}_t > n-k\}$: {\em The lost update can not be recovered from the block and thus the age $\delta_i$ continues to grow. Again, using Corollary \ref{cor:pe} and \ref{cor:BEP}, we can write}
\begin{align}
\label{P_EC}
P(\cC) = \pers\BEP_\MDS(n-1,k).    
\end{align}
\end{itemize}
% \textbullet{}~Event $\cA$: $\mathsf{E}_t_{\{i \mod k\}} = 0$:  \textit{$\delta_i$ drops to $\ell$ as it takes $\ell$ slots to transmit a packet. From \eqref{pe}}, 
% \begin{align}
% \label{P_EA}
% P(\cA) = 1-\pers.    
% \end{align}
% \textbullet{}~Event $\cB$: $\mathsf{E}_{\{i \mod k\}} = 1 ~\&~ \sum_{t=1}^n \mathsf{E}_t_t \leq n-k$: \textit{Since the number of erasures in the block is less than $n-k$, $\delta_i$ drops to $(n+1-\{i \mod k\})\ell$. From Section~\ref{CMBEP}}, 
% \begin{align}
% \label{P_EB}
% P(\cB) = \pers\left(1-\BEP_\MDS(n-1,k)\right).    
% \end{align}
% \textbullet{}~Event $\cC$: \textit{$\mathsf{E}_t_{\{i \mod k\}} = 1 ~\&~ \sum_{t=1}^n \mathsf{E}_t > n-k$. Since the block cannot recover the lost samples, they are discarded. From Section \ref{CMBEP},} 
% \begin{align}
% \label{P_EC}
% P(\cC) = \pers.\BEP_\MDS(n-1,k).    
% \end{align}

For $j\in \mathbb{Z}_{+}$, let  $\mathsf{R}_j$ describe how the $j$-th successful reception occurs such that $\mathsf{R}_j = 0$ correspond to Event $\cA$ and  $\mathsf{R}_j =1$ correspond to Event $\cB$. Since $\cA$ and $\cB$ are disjoint events, it follows that 
% \begin{align}
%     \label{P_R0}
%     P(\mathsf{R}_j =0) &= P(\cA|\cA,\cB) = \frac{P(\cA)}{P(\cA)+P(\cB)},\\
%    \text{and}~~ P(\mathsf{R}_j =1) &= P(\cB|\cA,\cB) = \frac{P(\cB)}{P(\cA)+P(\cB)}. 
%     \label{P_R1}
% \end{align}
\begin{align}
    \label{P_R0}
    P(\mathsf{R}_j =0) &=  \frac{P(\cA)}{P(\cA)+P(\cB)},\\
   \text{and}~~ P(\mathsf{R}_j =1) &=\frac{P(\cB)}{P(\cA)+P(\cB)}. 
    \label{P_R1}
\end{align}
Given the number of transmission between $(j-1)$-th and $j$-th receptions are $x$, $\mathsf{X}_j$  can take values equal to:\\
1) $\tilde{x}_1=x\lambda n\ell$ when $\mathsf{R}_{j} - \mathsf{R_{j-1}}=0$,\\
2) $\tilde{x}_2=x\lambda n\ell + (n-\{i \mod k\})\ell$ when $\mathsf{R}_{j} - \mathsf{R_{j-1}}=1$,\\
3) $\tilde{x}_3=x\lambda n\ell - (n-\{i \mod k\})\ell$ when  $\mathsf{R}_{j} - \mathsf{R_{j-1}}=-1$. \newline
Thus, using \eqref{P_EA}-\eqref{P_R1}, the pmf of $\mathsf{X}_j$ can be derived as
% \begin{align*}
% P\big(\mathsf{X}_j &= x\lambda n\ell\big) = P(\cC)^{x-1}(1-P(\cC))P(\mathsf{R}_{j} - \mathsf{R_{j-1}} = 0)\\
% &= P(\cC)^{x-1}\left(\frac{P(\cA)^2 + P(\cB)^2}{P(\cA)+P(\cB)}\right).\\
% P\big(\mathsf{X}_j &= x\lambda n\ell + (n-\{i \mod k\})\ell\big) \\
% &= P(\cC)^{x-1}(1-P(\cC))P(\mathsf{R}_{j} - \mathsf{R_{j-1}} = 1)\\
% &= P(\cC)^{x-1}\left(\frac{P(\cB).P(\cA)}{P(\cA)+P(\cB)}\right).\\
% P\big(\mathsf{X}_j &= x\lambda n\ell - (n-\{i \mod k\})\ell\big) \\
% &= P(\cC)^{x-1}(1-P(\cC))P(\mathsf{R}_{j} - \mathsf{R_{j-1}} = -1)\\
% &= P(\cC)^{x-1}\left(\frac{P(\cA).P(\cB)}{P(\cA)+P(\cB)}\right).
% \end{align*}
% \begin{align*}
% P\big(\mathsf{X}_j &= \tilde{x}_1\big) = P(\cC)^{x-1}\left(\frac{P(\cA)^2 + P(\cB)^2}{P(\cA)+P(\cB)}\right),\\
% P\big(\mathsf{X}_j &= \tilde{x}_m\big) = P(\cC)^{x-1}\left(\frac{P(\cB)P(\cA)}{P(\cA)+P(\cB)}\right),
% \end{align*}
% for $m\in\{\2,3\}$
% for $m\in\{2,3\}$.
\begin{align*}
P\big(\mathsf{X}_j = \tilde{x}_m\big) =\begin{cases}
    P(\cC)^{x-1}\left(\frac{P(\cA)^2 + P(\cB)^2}{P(\cA)+P(\cB)}\right)&\text{~~for~}m=1,\\
 P(\cC)^{x-1}\left(\frac{P(\cB)P(\cA)}{P(\cA)+P(\cB)}\right)&\text{~~for~}m=2,3.
\end{cases} 
\end{align*}
Using this pmf and simple algebraic calculations, we can derive the first two moments of $\mathsf{X}_j$ as follows 
\begin{align}
\label{Ex_cod}
&\mathbb{E}[\mathsf{X}_j]   = \frac{\lambda n \ell}{1-P(\cC)} \text{~~and~~} \\
\label{Ex2_cod}
&\mathbb{E}[\mathsf{X}_j^2]=\frac{(1+P(\cC))(\lambda n \ell)^2 + 2P(\cA)P(\cB)\big((n-\{i \mod k\})\ell\big)^{2}}{(1-P(\cC))^2}.
\end{align}
% \begin{align}
% \label{Ex_cod}
% \mathbb{E}[\mathsf{X}_j]  =  \sum_{x=1}^{\infty} x\lambda n \ell \big(P(\cA) + P(\cB)\big)P(\cC)^{x-1}  = \frac{\lambda n \ell}{1-P(\cC)},
% \end{align}
% \begin{align}
% \label{Ex2_cod}
% \begin{split}
% \mathbb{E}[\mathsf{X}_j^2] &= \sum_{x=1}^{\infty} (x\lambda n \ell )^2\big(P(\cA) + P(\cB)\big)P(\cC)^{x-1}\\
% &+ \sum_{x=1}^{\infty} \frac{2P(\cA)P(\cB)}{P(\cA)+P(\cB)}\big((n-\{i \mod k\})\ell\big)^{2} P(\cC)^{x-1}.\\
% &= \frac{(1+P(\cC))(\lambda n \ell)^2 + 2P(\cA)P(\cB)\big((n-\{i \mod k\})\ell\big)^{2}}{(1-P(\cC))^2}.
% \end{split}
% \end{align}
Using \eqref{eq:AvgAoI}, \eqref{Ex_cod} and \eqref{Ex2_cod}, we obtain average $\AoI$ in the following theorem.
\begin{theorem}
\label{lemma:avgAoI_MDS}
    The average $\AoI$ of source $S_i$ for  $(n,k)-\MDS$ coded update system is 
\begin{align}
    \Delta_i = \frac{\lambda n \ell(1+P(\cC))}{2(1-P(\cC))} + \frac{P(\cA)P(\cB)\big((n-\{i \mod k\})\ell\big)^{2}}{\lambda n \ell(1-P(\cC))} + \ell.
\end{align}
If $K\gg\max\left\{\sqrt{\frac{2P(\cA)P(\cB)\big((n-\{i \mod k\})\big)^{2}}{(1+P(\cC))}}, \frac{2(1-P(\cC))}{1+P(\cC)}\right\}$, the average $\AoI$ for a generic source  can be approximated as
$$\Delta_{\MDS}(k,n) \approx\dfrac{K n\ell(1+P(\cC))}{2k(1-P(\cC))}.$$ 
\end{theorem}
\begin{figure*}[h]
    \centering
    \includegraphics[width=0.55\textwidth]{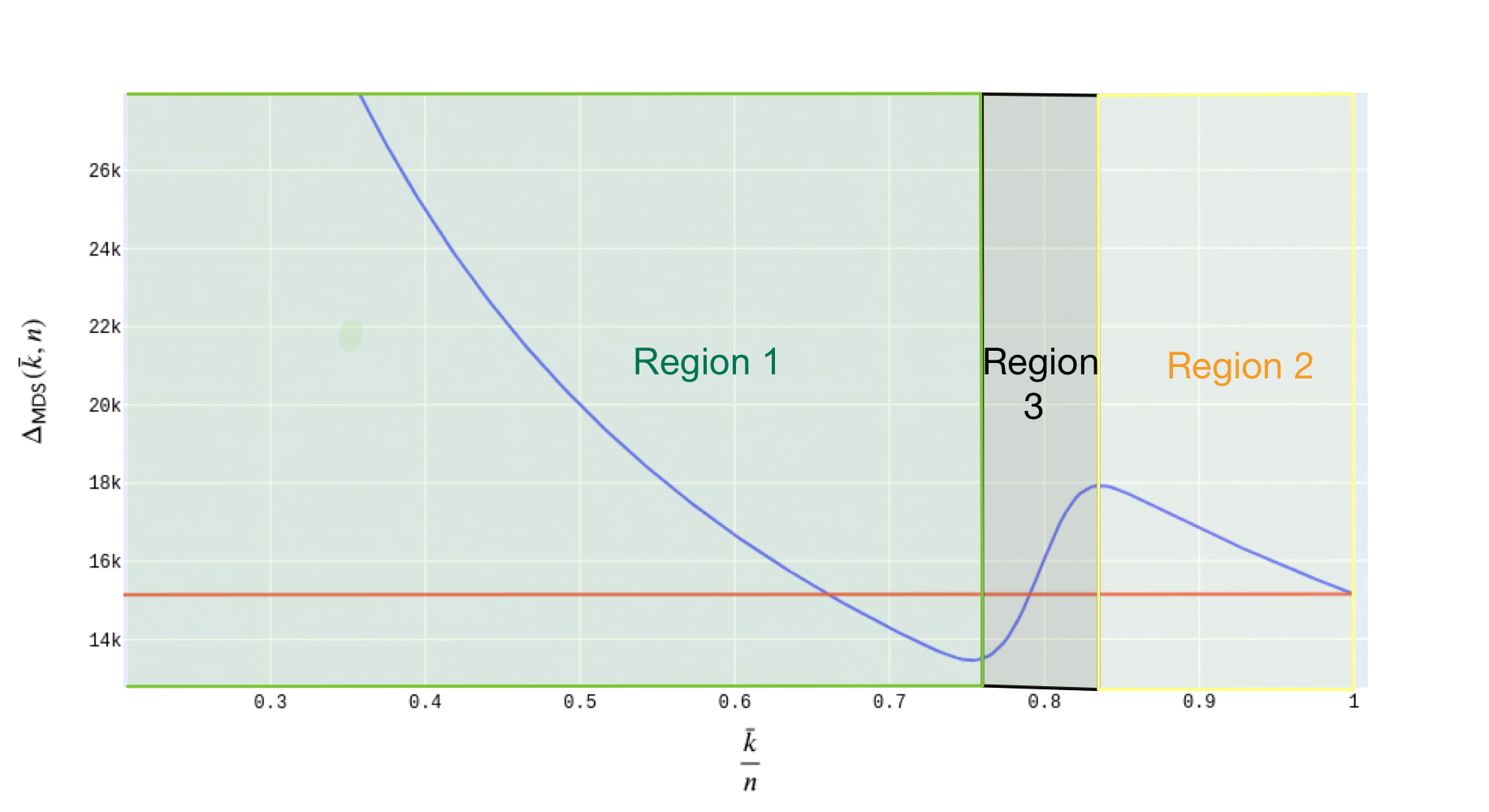}\vspace{-2mm}
    \caption{$\Delta_{\MDS}(\bar{k},n)$ v/s $\frac{\bar{k}}{n}$ for $n = 300, K=10000, \alpha = 0.2, \beta = 0.8, \epsilon_0 = 0.2, \epsilon_1 = 0.9$. Here, the red line corresponds to the uncoded system and the blue line to the coded system.}% \textcolor{red}{Indicate the regions, increase x-axis and y-axis labels font size.} \textcolor{blue}{Are we including curve for one more value of $n$?}
    \label{fig:comp}\vspace{-3mm}
\end{figure*}

In the next subsection, we analyse $\Delta_{\MDS}(k,n)$ to find the optimal value of block size $k$ for a given size of the coded block $n$ and given channel channel parameters ($\alpha, \beta, \epsilon_0, \epsilon_1$).
\vspace{-0.3cm}
\subsection{Analysis of $\Delta_{\MDS}(k,n)$}
In this subsection, we focus on finding the optimal $k$ which minimizes the average $\AoI$ derived in Theorem \ref{lemma:avgAoI_MDS}.
Let $\mathscr{K}(n)$ be defined as 
$$\mathscr{K}(n) \triangleq  \underset{1\leq k\leq n}{\argmin}~\Delta_{\MDS}(k,n).
$$
Recall the random variables $\mathsf{E}_{t}$ that indicate the presence of an erasure at time $t$. By central limit theorem  for Markov chains \cite{geyer1998markov}, we can write 
\begin{align*}
    \sum_{t=1}^{n} \mathsf{E}_t 
    \overset{\text{dist}}{\longrightarrow} \cN\big(n\mathbb{E}[\mathsf{E}_{1}], n\sigma^2\big),
\end{align*}  
when $n$ is large, 
where $\sigma^2 = \text{Var}(\mathsf{E}_1) + 2\sum_{t=1}^{n} \text{Cov}(\mathsf{E}_1,\mathsf{E}_{1+t})$. Since  $\mathsf{E}_i$ and $\mathsf{E}_j$ are indicator variables, we have  $\mathbb{E}[\mathsf{E}_{i}\mathsf{E}_j] = \mathbb{E}[\mathsf{E}_{i}]\mathbb{E}[\mathsf{E}_{j}]$ from which we get $\text{Cov}(\mathsf{E}_1,\mathsf{E}_{1+t}) = 0$. Hence, $\sigma^2 = \text{Var}(\mathsf{E}_1) = \pers(1-\pers)$. Therefore, we have 
\begin{align*}
    \sum_{t=1}^{n} \mathsf{E}_t 
    \overset{\text{dist}}{\longrightarrow} \cN\big(n\pers, n\pers(1-\pers)\big).
\end{align*}
From the cumulative distribution function (CDF) $\Phi(\cdot)$ of a standard normal distribution $\cN(0,1)$, we can write
\begin{align*}
    P\bigg(\sum_{t=1}^{n} \mathsf{E}_t &>n\cP - c_n\sqrt{n\cP(1-\cP)}\bigg)=\Phi(c_n), 
\end{align*}
where $c_n \in \mathbb{R}$. Thus, for $\bar{k} \triangleq n-n\pers+c_n\sqrt{n\pers(1-\pers)}$, the probaility of Event $\cC$ given in \eqref{P_EC} can be written as $P(\cC)=\pers\Phi(c_n)$ and thus the average $\AoI$ given in Theorem \ref{lemma:avgAoI_MDS} can be calculated as given in next lemma.
\begin{lemma}
The average $\AoI$ for a generic source  $(n,\bar{k})-\MDS$ coded update system  can be approximated as  $\Delta_{\MDS}(\bar{k},n) = $ 
\begin{align}
\label{Delta_k'}
\dfrac{K\ell\big(1+\pers\Phi(c_n)\big)}{2\left(1-\pers+c_n\sqrt{\frac{\pers(1-\pers)}{n}}\right)\big(1-\pers\Phi(c_n)\big)}.
\end{align}
\end{lemma}
\noindent In Fig.~\ref{fig:comp}, we illustrate the behavior of $\Delta_{\MDS}(\bar{k},n)$ given by \eqref{Delta_k'} with respect to the coding rate $R \triangleq{\bar{k}}/{n}$.
%In Fig. \ref{fig:comp}, we simulate \eqref{Delta_k'} to analyse the behavior of $\Delta_{\MDS}(\bar{k},n)$ w.r.t. $c_n$.
Let $c_{\epsilon} \in \mathbb{R_{+}}$ be such that $\frac{e^{-c_n^2}}{|c_n|} < \epsilon$   for $|c_n| > c_{\epsilon}$. It can be verified that $c_{\epsilon} \sim \Theta\left(\sqrt{\ln\left(\frac{1}{\epsilon}\right)}\right)$ suffices. Further, it is to be noted that $\Phi(c_n)$ quickly approaches to zero and one when  $c_n<-c_\epsilon$ and $c_n>c_\epsilon$, respectively. Using this, in the following, we analyze the age over different regions (shown in Fig.~\ref{fig:comp}).
\newline {\em Region 1}: We have $\Phi(c_n) \sim \cO\left(\frac{e^{-c_n^2}}{|c_n|}\right)$ when $c_n<-c_\epsilon$. Since $\Phi(c_n)$ decreases exponentially with decrease in $c_n$, we can rewrite  \eqref{Delta_k'} as
\begin{align}
\Delta_{\MDS}(\bar{k},n) &= \dfrac{K\ell}{2\left(1-\pers+c_n\sqrt{\frac{\pers(1-\pers)}{n}}\right)},
\label{Region1}
\end{align}
for $-\sqrt{\frac{n(1-\pers)}{\pers}}<c_n< -c_{\epsilon}$.
\newline {\em Region 2}: We also have $\big(1-\Phi(c_n)\big) \sim \cO\left(\frac{e^{-c_n^2}}{|c_n|}\right)$ for $c_n>c_\epsilon$. Besides, $1-\Phi(c_n)$ decays exponentially to zero with increasing $c_n$. 
Hence, we can rewrite \eqref{Delta_k'} as
\begin{align}
    \Delta_{\MDS}(\bar{k},n) &= \dfrac{K\ell\left(1+\pers\right)}{2\left(1-\pers+c_n\sqrt{\frac{\pers(1-\pers)}{n}}\right)\left(1-\pers\right)}.\label{Region2}
\end{align}
for $c_\epsilon<c_n< \sqrt{\frac{n\pers}{1-\pers}}$.
\newline {\em Region 3}: From $\Phi(c_n) \sim \cO\left(\frac{e^{-c_n^2}}{|c_n|}\right)$, it is easy to deduce that $\Phi(c_n)$ is approximately linear over $[-c_\epsilon,c_\epsilon]$. In fact, we can show that $\Phi(c_n) \approx \frac{c_n+c_\epsilon}{2c_\epsilon}$ for $-c_\epsilon < c_n < c_\epsilon$. Using this,   we can rewrite \eqref{Delta_k'} as
\begin{align}
\Delta_{\MDS}(\bar{k},n) &= \dfrac{K\ell\big(1+\pers\frac{c_n+c_\epsilon}{2c_\epsilon}\big)}{2\left(1-\pers\right)\big(1-\pers\frac{c_n+c_\epsilon}{2c_\epsilon}\big)}.\label{Region3}
\end{align}
for $-c_\epsilon < c_n < c_\epsilon$. 

From \eqref{Region2} and \eqref{Region3}, it is evident that the local maxima is  $\Delta^{\star\star}_{\MDS}(\bar{k},n) =\dfrac{K\ell\big(1+\pers\big)}{2\left(1-\pers+\smallO(1)\right)(1-\pers)}$ at $\bar{k}=n(1-\pers) + \smallO(n)$.
Finally, we  obtain the optimal average $\AoI$ in the following theorem by  selecting $\bar{k}$ or $c_n$ that minimizes \eqref{Delta_k'} (or maximizes the denominator of \eqref{Region1}).
\begin{theorem}
For a given blocklength $n$, the  optimal $\mathscr{K}(n)$  that minimizes the average $\AoI$ of $(n,k)-\MDS$ coded update system is given by   \begin{align}
\mathscr{K}(n) = n(1-\pers) - \smallO(n),    
\end{align} 
and the minimum average $\AoI$ is 
\begin{align}
\Delta_{\MDS}^\star(\bar{k},n) &= \dfrac{K\ell}{2\left(1-\pers+\smallO(1)\right)}.
\label{eq:optimalAge_MDScodes}
\end{align}
\end{theorem}
Comparing the minimum $\AoI$ given in \eqref{eq:optimalAge_MDScodes} of coded system  with the $\AoI$ of undocded system given in Lemma \ref{lemma:Age_uncoded}, we determine the maximum coding gain in the following corollary.
\begin{corollary}
    The maximum coding gain for the average $\AoI$ of $(n,k)-\MDS$ coded update system is given by
    \begin{align}
        \mathscr{G}_c=\frac{\Delta^\star_{\MDS}(\bar{k},n)}{\Delta_{\mathsf{Uncoded}}}=1+\pers-\smallO(1).
    \end{align}
\end{corollary}

% \bibliographystyle{IEEEtran} 
% \bibliography{output.bbl}

\begin{thebibliography}{10}
\bibitem{Jiang}
W.~Jiang, B.~Han, M.~Habibi, and H.~Schotten, ``The road towards {6G}: A
  comprehensive survey,'' \emph{IEEE Open J. Commun. Soc.}, vol.~2, pp.
  334--366, 2021.

\bibitem{RoyYates_2021Survey}
R.~D.~Yates, Y.~Sun, D.~R.~Brown, S.~K.~Kaul, E.~Modiano, and S.~Ulukus, ``Age
  of information: An introduction and survey,'' \emph{IEEE J. Sel. Areas
  Commun.}, vol.~39, no.~5, pp. 1183--1210, 2021.

\bibitem{ChenKun_2016}
K.~Chen and L.~Huang, ``Age-of-information in the presence of error,'' in
  \emph{IEEE ISIT}, 2016, pp. 2579--2583.

\bibitem{FengSongtao_2019RatelessCode_OptimalAoI}
S.~Feng and J.~Yang, ``Age-optimal transmission of rateless codes in an erasure
  channel,'' in \emph{IEEE ICC}, 2019, pp. 1--6.

\bibitem{RoyYates_2017}
R.~D.~Yates, E.~Najm, E.~Soljanin, and J.~Zhong, ``Timely updates over an
  erasure channel,'' in \emph{IEEE ISIT}.\hskip 1em plus 0.5em minus
  0.4em\relax IEEE, 2017, pp. 316--320.

\bibitem{Xie_2020ARQ}
M.~Xie, Q.~Wang, J.~Gong, and X.~Ma, ``Age and energy analysis for {LDPC} coded
  status update with and without {ARQ},'' \emph{IEEE Int. Things J.}, vol.~7,
  no.~10, pp. 10\,388--10\,400, 2020.

\bibitem{ParagParimal_2017BEC}
P.~Parag, A.~Taghavi, and J.-F.~Chamberland, ``On real-time status updates over
  symbol erasure channels,'' in \emph{IEEE WCNC}, 2017, pp. 1--6.

\bibitem{ElieNajm_2017HARQ}
E.~Najm, R.~Yates, and E.~Soljanin, ``Status updates through {M/G/1/1} queues
  with {HARQ},'' in \emph{IEEE ISIT}, 2017, pp. 131--135.

\bibitem{ElieNajm_2019OptimalAge}
E.~Najm, E.~Telatar, and R.~Nasser, ``Optimal age over erasure channels,'' in
  \emph{IEEE ISIT}, 2019, pp. 335--339.

\bibitem{FengSongtao_2019Conding_BroadcastNetwork}
S.~Feng and J.~Yang, ``Adaptive coding for information freshness in a two-user
  broadcast erasure channel,'' in \emph{IEEE GLOBECOM}, 2019, pp. 1--6.

\bibitem{Xingran_2019Conding_BroadcastNetwork}
X.~Chen and S.~S.~Bidokhti, ``Benefits of coding on age of information in
  broadcast networks,'' in \emph{IEEE ITW}, 2019, pp. 1--5.

\bibitem{FaraziShahab_2020}
S.~Farazi, A.~G.~Klein, and D.~R.~Brown, ``Average age of information in update
  systems with active sources and packet delivery errors,'' \emph{IEEE Wireless
  Commun. Lett.}, vol.~9, no.~8, pp. 1164--1168, 2020.


\bibitem{gilbert}
E.~N. Gilbert, ``Capacity of a burst-noise channel,'' \emph{Bell system
  technical journal}, vol.~39, no.~5, pp. 1253--1265, 1960.

\bibitem{elliott}
E.~O. Elliott, ``Estimates of error rates for codes on burst-noise channels,''
  \emph{The Bell System Technical Journal}, vol.~42, no.~5, pp. 1977--1997,
  1963.

\bibitem{HasHoh}
G.~Ha{\ss}linger and O.~Hohlfeld, ``The {Gilbert-Elliott} model for packet loss
  in real time services on the internet,'' in \emph{Proc. 14th {GI/ITG}
  Conference on Measurement, Modelling and Evaluation of Computer and
  Communication Systems}.\hskip 1em plus 0.5em minus 0.4em\relax {VDE} Verlag,
  2008, pp. 269--286.

  
\bibitem{MartTrotISIT07}
E.~Martinian and M.~Trott, ``{Delay-Optimal Burst Erasure Code Construction},''
  in \emph{{Proc. Int. Symp. Inf. Theory, Nice, France, June 24-29,
  2007}}.\hskip 1em plus 0.5em minus 0.4em\relax {IEEE}, pp. 1006--1010.

\bibitem{BadrPatilKhistiTIT17}
A.~Badr, P.~Patil, A.~Khisti, W.~Tan, and J.~G.~Apostolopoulos, ``{Layered
  Constructions for Low-Delay Streaming Codes},'' \emph{{IEEE} Trans. Inf.
  Theory}, vol.~63, no.~1, pp. 111--141, 2017.

\bibitem{NikDeepPVK}
M.~N.~Krishnan, D.~Shukla, and P.~V.~Kumar, ``{Low Field-size, Rate-Optimal
  Streaming Codes for Channels With Burst and Random Erasures},'' 2019,
  arXiv:1903.06210.


\bibitem{NikRamVajKum}
M.~N.~Krishnan, V.~Ramkumar, M.~Vajha, and P.~V.~Kumar, ``Simple streaming
  codes for reliable, low-latency communication,'' \emph{{IEEE} Commun. Lett.},
  vol.~24, no.~2, pp. 249--253, 2020.

\bibitem{SinRKum22}
S.~Singhvi, R.~Gayathri and P.~V.~Kumar, "Rate-Optimal Streaming Codes Over the Three-Node Decode-And-Forward Relay Network," 2022 IEEE International Symposium on Information Theory (ISIT), Espoo, Finland, 2022, pp. 1957-1962, doi: 10.1109/ISIT50566.2022.9834645.


\bibitem{ShobRamKum22}
S.~Bhatnagar, V.~Ramkumar and P.~V.~Kumar, "Rate-Optimal Streaming Codes with Smaller Field Size Under Less-Stringent Decoding-Delay Requirements," 2022 IEEE Information Theory Workshop (ITW), Mumbai, India, 2022, pp. 612-617, doi: 10.1109/ITW54588.2022.9965940.


\bibitem{RudowRashmi22}
M.~Rudow and K.~V.~Rashmi, "Learning-Augmented Streaming Codes are Approximately Optimal for Variable-Size Messages," 2022 IEEE International Symposium on Information Theory (ISIT), Espoo, Finland, 2022, pp. 474-479, doi: 10.1109/ISIT50566.2022.9834539.

\bibitem{HannaAnt23}
Hanna, S. K., Tan, Z., Xu, W., Wachter-Zeh, A. (2023). Codes Correcting Burst and Arbitrary Erasures for Reliable and Low-Latency Communication. arXiv preprint arXiv:2302.08644.

\bibitem{VajRamJhaKum20}
M.~Vajha, V.~Ramkumar, M.~Jhamtani, and P.~V.~Kumar, ``On sliding window
  approximation of gilbert-elliott channel for delay constrained setting,''
  \emph{CoRR}, vol. abs/2005.06921, 2020.
  
\bibitem{VajRamJhaKum21}
M.~Vajha, V.~Ramkumar, M.~Jhamtani, and P.~V.~Kumar, ``On the performance
  analysis of streaming codes over the {Gilbert-Elliott} channel,'' in
  \emph{IEEE ITW}, 2021, p. 1–6.

\bibitem{geyer1998markov}
C.~J.~Geyer, ``Markov chain monte carlo lecture notes,'' \emph{Course notes,
  Spring Quarter}, vol.~80, 1998.

\end{thebibliography}

\newpage
\input{ref.bbl}
\end{document}